\documentclass[runningheads]{llncs}

\title{Who Asked \emph{Us?}\\How the Theory of Computing Answers Questions about Analysis}
\subtitle{\vspace{1em}\normalfont{\emph{Dedicated to the Memory of Ker-I Ko}}}
\titlerunning{Who Asked \emph{Us?}}

\author{Jack H. Lutz\thanks{Research supported in part by National Science Foundation grants 1545028 and 1900716.} \and Neil Lutz}
\institute{Iowa State University (\email{\{lutz,nlutz\}@iastate.edu})}

\usepackage{amsmath}
\usepackage{amssymb}
\usepackage{mathtools}
\usepackage{url}
\usepackage[hidelinks]{hyperref}

\numberwithin{equation}{section}

\DeclareMathOperator{\proj}{proj}

\DeclareMathOperator{\Dim}{Dim}
\DeclareMathOperator{\dimH}{dim_H}

\DeclareMathOperator{\dimP}{dim_P}

\newcommand{\R}{\mathbb{R}}

\newcommand{\N}{\mathbb{N}}
\newcommand{\Q}{\mathbb{Q}}

\begin{document}

\maketitle

\begin{abstract}
	Algorithmic fractal dimensions---constructs of computability theory---have recently been used to answer open questions in \emph{classical} geometric measure theory, questions of mathematical analysis whose statements do not involve computability theory or logic. We survey these developments and the prospects for future such results.
\end{abstract}

\section{Introduction}

Ker-I Ko was a pioneer in the computability, and especially the computational complexity, of problems in mathematical analysis. Aside from his visionary work on the complexity theory of functions on the reals, the early part of which is summarized in his well-known 1991 monograph~\cite{Ko91}, he did groundbreaking work on computability and complexity aspects of fractal geometry and other topics in geometric measure theory~\cite{Ko95,ChouK95,KoW96,Ko98,ChouK04,ChouK05a,ChouK05b,YuCK06,KoY07,Ko13}. 

This chapter surveys recent developments in which algorithmic fractal dimensions, which are constructs of the theory of computing, have been used to answer open questions in {\it classical} fractal geometry, questions of mathematical analysis whose statements do not involve the theory of computing.

The results surveyed here concern the classical Hausdorff and packing dimensions of sets in Euclidean spaces $\R^n$. These fractal dimensions are duals of each other that were developed in 1918 and the early 1980s, respectively~\cite{Haus18,Tric82,Sull84}. They assign every set $E \subseteq \R^n$ a \emph{Hausorff dimension} $\dimH(E)$ and a \emph{packing dimension} $\dimP(E)$, which are real numbers satisfying $0 \leq \dimH(E)\leq \dimP(E) \leq n$~\cite{Falc14,Edga08}. These dimensions are both 0 if $E$ consists of a single point, 1 if $E$ is a smooth curve, $2$ if E is a smooth surface, etc., but, for any two real numbers $\alpha$ and $\beta$ satisfying $0 \leq \alpha \leq \beta \leq n$, there are $2^\mathfrak{c}$ many sets $E \subseteq \R^n$ such that $\dimH(E) = \alpha$ and $\dimP(E) = \beta$, where $\mathfrak{c} = 2^{\aleph_0}$ is the cardinality of the continuum. So-called ``fractals'' (a term with no accepted formal definition) are typically sets $E \subseteq \R^n$ with non-integral Hausdorff and packing dimensions. (Note: Hausdorff and packing dimensions are well-defined in arbitrary metric spaces, but this generality is not needed in the present survey.)

In contrast with the above classical fractal dimensions, the algorithmic fractal dimensions developed in~\cite{Lutz03b,AHLM07} and defined in Section~\ref{sec:aiad} below use computability theory to assign each \emph{individual point} $x$ in a Euclidean space $\R^n$ a \emph{dimension} $\dim(x)$ and a \emph{strong dimension} $\Dim(x)$ satisfying $0 \leq \dim(x) \leq \Dim(x) \leq n$. Intuitively, $\dim(x)$ and $\Dim(x)$ are the lower and upper densities of the algorithmic information in $x$. Computable points $x$ (and many other points) satisfy $\dim(x) = \Dim(x) = 0$. In contrast, points $x$ that are algorithmically random in the sense of Martin-L\"{o}f~\cite{Mart66} (and many other points) satisfy $\dim(x) = \Dim(x) = n$. In general, for any two real numbers $\alpha$ and $\beta$ satisfying $0 \leq \alpha \leq \beta \leq n$, the set of points $x \in \R^n$ such that $\dim(x) = \alpha$ and $\Dim(x) = \beta$ has the cardinality $\mathfrak{c}$ of the continuum.

The algorithmic fractal dimensions $\dim(x)$ and $\Dim(x)$ were known from their inceptions to be closely related to---and in fact $\Sigma^0_1$ versions of---their respective classical forerunners $\dimH(E)$ and $\dimP(E)$~\cite{Lutz03b,AHLM07}. However, it was only recently~\cite{LutLut18} that the \emph{point-to-set principles} discussed in section 3 below were proven, giving complete characterizations of $\dimH(E)$ and $\dimP(E)$ in terms of oracle relativizations of $\dim(x)$ and $\Dim(x)$, respectively. 

The point-to-set principles are so named because they enable one to infer a bound---especially a difficult lower bound---on the classical fractal dimensions of a set $E \subseteq \R^n$ from a bound on the relativized algorithmic dimension of a single, judiciously chosen point $x \in E$. The power of this point-to-set reasoning has quickly become apparent. Sections~\ref{sec:prod} through~\ref{sec:proj} below survey recent research in which this method has been used to prove new theorems in classical fractal geometry. Several of these theorems answered well-known open questions in the field, completely classical questions whose statements do not involve computability or logic. Section~\ref{sec:conc} discusses the prospects for future such results.

\section{Algorithmic Information and Algorithmic Dimensions}\label{sec:aiad}

The \emph{Kolmogorov complexity}, or \emph{algorithmic information content}, of a string $x\in\{0,1\}^*$ is
\[K(x)=\min\big\{|\pi|\;\big|\;\pi\in\{0,1\}^*\text{ and }U(\pi)=x\big\}\,,\]
where $U$ is a fixed universal prefix Turing machine, and $\pi$ is the length of a binary ``program $\pi$ for $x$.'' Extensive discussions of the history and intuition behind this notion, including its essential invariance with respect to the choice of the universal Turing machine $U$, may be found in any of the standard texts~\cite{LiVit08,DowHir10,Nies09,ShUsVe17}. By routine encoding we extend this notion to let $x$ range over various countable sets, so that $K(x)$ is well defined when $x$ is an element of $\N$, $\Q$, $\Q^n$, etc.

We ``lift'' Kolmogorov complexity to Euclidean space in two steps. We first define the Kolmogorov complexity of a set $E\subseteq\R^n$ to be
\[K(E)=\min\{K(q)\mid q\in\Q^n\cap E\}\,,\]
i.e., the amount of information to specify \emph{some} rational point in $E$. (A similar notion was used for a different purpose in~\cite{SheVer02}.) Note that
\[E\subseteq F\implies K(E)\geq K(F)\,.\]
We then define the \emph{Kolmogorov complexity} of a point $x\in\R^n$ at a \emph{precision} $r\in\N$ to be
\[K_r(x)=K\big(B_{2^{-r}}(x)\big)\,,\]
where $B_\varepsilon(x)$ is the open ball of radius $\varepsilon$ about $x$. That is, $K_r(x)$ is the number of bits required to specify \emph{some} rational point $q$ whose Euclidean distance from $x$ is less than $2^{-r}$.

The \emph{(algorithmic) dimension} of a point $x\in\R^n$ is
\begin{equation}\label{eq:dimdef}
\dim(x)=\liminf_{r\to\infty}\frac{K_r(x)}{r}\,,
\end{equation}
and the \emph{strong (algorithmic) dimension} of a point $x\in\R^n$ is
\begin{equation}\label{eq:Dimdef}
\Dim(x)=\limsup_{r\to\infty}\frac{K_r(x)}{r}\,.
\end{equation}
(The adjectives ``constructive'' and ``effective'' are sometimes used in place of ``algorithmic'' here.) We should note that the identities~(\ref{eq:dimdef}) and~(\ref{eq:Dimdef}) were originally \emph{theorems} proven in~\cite{LutMay08} (following a key breakthrough in~\cite{Mayo02}) characterizing the algorithmic dimensions $\dim(x)$ and $\Dim(x)$ that had first been developed using algorithmic betting strategies called gales~\cite{Lutz03b,AHLM07}. The characterizations~(\ref{eq:dimdef}) and~(\ref{eq:Dimdef}) support the intuition that $\dim(x)$ and $\Dim(x)$ are the lower and upper asymptotic densities of algorithmic information in the point $x\in\R^n$.

By giving the underlying universal prefix Turing machine oracle access to a set $A\subseteq\N$, the quantities in this section can all be defined \emph{relative} to $A$. We denote these relativized complexities and dimensions by $K^A(x)$, $K^A_r(x)$, $\dim^A(x)$, etc. When $A$ encodes a point $y\in\R^n$, we may instead write $K^y(x)$, $K^y_r(x)$, $\dim^y(x)$, etc. The following easily verified result is frequently useful.
\begin{theorem}[chain rule for algorithmic dimensions]\label{thm:chainrule}
	For all $x\in\R^m$ and $y\in\R^n$,
	\begin{align*}
	\dim^y(x)+\dim(y)&\leq\dim(x,y)\\
	&\leq\Dim^y(x)+\dim(y)\\
	&\leq\Dim(x,y)\\
	&\leq\Dim^y(x)+\Dim(y)\,.
	\end{align*}
\end{theorem}

\section{Point-to-Set Principles}\label{sec:p2s}

One of the oldest and most beautiful theorems of computable analysis says that a function $f:\R\to \R$ is continuous if and only if there is an oracle $A\subseteq\N$ relative to which $f$ is computable~\cite{Mosc80,Soar09}. That is, relativization allows us to characterize continuity---a completely classical notion---in terms of computability. The following two recent theorems are very much in the spirit of this old theorem.

\begin{theorem}[point-to-set principle for Hausdorff dimension~\cite{LutLut18}]\label{thm:hp2s}
	For every set $E\subseteq\R^n$,
	\begin{equation}\label{eq:hp2s}
		\dimH(E)=\adjustlimits\min_{A\subseteq\N}\sup_{{x}\in E}\,\dim^A({x})\,.
	\end{equation}
\end{theorem}
\begin{theorem}[point-to-set principle for packing dimension~\cite{LutLut18}]\label{thm:pp2s}
	For every set $E\subseteq\R^n$,
	\begin{equation}\label{eq:pp2s}
		\dimP(E)=\adjustlimits\min_{A\subseteq\N}\sup_{{x}\in E}\,\Dim^A({x})\,.
	\end{equation}
\end{theorem}

For purposes of this survey, readers unfamiliar with Hausdorff and packing dimensions may use Theorems~\ref{thm:hp2s} and~\ref{thm:pp2s} as their \emph{definitions}, but it should be kept in mind that these characterizations are theorems that were proven a century after Hausdorff developed his beautiful dimension.

Two remarks on the point-to-set principles are in order here. First, as the principles state, the minima on the right-hand sides of~(\ref{eq:hp2s}) and~(\ref{eq:pp2s}) are actually achieved. In other words, if we define a \emph{Hausdorff oracle} for a set $E \subseteq \R^n$ to be an oracle $A \subseteq \N$ such that
\begin{equation}\label{eq:horacle}
	\dimH(E)=\sup_{{x}\in E}\,\dim^A({x})\,,
\end{equation}
and we similarly define a \emph{packing oracle} for a set $E \subseteq \R^n$ to be an oracle $A \subseteq \N$ such that
\begin{equation}\label{eq:poracle}
	\dimP(E)=\sup_{{x}\in E}\,\Dim^A({x})\,,
\end{equation}
then the point-to-set principles are assertions that \emph{every} set $E \subseteq \R^n$ has Hausdorff and packing oracles. It is easy to show that, if $A$ is a Hausdorff oracle for a set $E \subseteq \R^n$, and if $A$ is Turing reducible to a set $B \subseteq \N$, then $B$ is also a Hausdorff oracle for $E$, and similarly for packing oracles. This is useful, because it often enables one to combine Hausdorff or packing oracles with other oracles in a proof.

The second remark on the point-to-set principles concerns their use. Some of the most challenging problems in fractal geometry and dynamical systems involve finding lower bounds on the fractal dimensions of various sets. The point-to-set principles allow us to infer lower bounds on the fractal dimensions of sets from lower bounds on the corresponding relativized algorithmic fractal dimensions of judiciously chosen \emph{individual points} in those sets. For example, to prove, for a given set $E \subseteq \R^n$, that $\dimH(E) \geq \alpha$, it suffices to show that, for every Hausdorff oracle $A$ for $E$ and every $\varepsilon > 0$, there is a point $x \in E$ such that $\dim^A(x) > \alpha - \varepsilon$. In some applications, the $\varepsilon$ here is not even needed, because one can readily show that there is a point $x \in E$ such that $\dim^A(x) \geq \alpha$. Most of the rest of this survey is devoted to illustrating the power of this point-to-set reasoning about fractal dimensions.

\section{Fractal Products}\label{sec:prod}

		Marstrand's product formula~\cite{Mars54,Falc14} states that for all sets $E,F\subseteq\R^n$,
		\[\dimH(E)\leq \dimH(E\times F)-\dimH(F)\,.\]
		The proof of this fact for Borel sets is simple~\cite{Falc14}, but Marstrand's original proof of the general result is more difficult~\cite{Matt95}. Using the point-to-set principle for Hausdorff dimension, the general result is an almost trivial consequence of the chain rule, Theorem~\ref{thm:chainrule}~\cite{Lutz17}. Tricot~\cite{Tric82} proved related inequalities about packing dimension, including the fact that for all $E,F\subseteq\R^n$,
		\[\dimP(E)\geq \dimH(E\times F)-\dimH(F).\]
		Xiao~\cite{Xiao96} showed that for every Borel set $E\subseteq\R^n$ and $\varepsilon>0$, there exists a Borel set $F\subseteq\R^n$ such that
		\begin{equation}\label{eq:bpx}
			\dimP(E)\leq \dimH(E\times F)-\dimH(F)+\varepsilon\,.
		\end{equation}
		Bishop and Peres~\cite{BisPer96} independently showed that for Borel (or analytic) $E$ there exists a compact $F$ satisfying~(\ref{eq:bpx}); they also later commented that that it would be straightforward to modify their construction to achieve $\varepsilon=0$.

		Using the point-to-set principles, N. Lutz proved for arbitrary sets $E$ that $\varepsilon=0$ can be achieved in~(\ref{eq:bpx}), albeit not necessarily by a compact or Borel set $F$.
		\begin{theorem}[\cite{Lutz19}]\label{thm:prod}
			For every set $E\subseteq\R^n$,
			\[\dimP(E)=\max_{F\subseteq\R^n}\big(\dimH(E\times F)-\dimH(F)\big)\,.\]
		\end{theorem}
		The particular set $F$ constructed in the proof of this theorem is the set of all points $x\in \R^n$ with $\dim^A(x)\leq n-\dimP(E)$, for a carefully chosen oracle $A$.

\section{Fractal Intersections}\label{sec:int}
	Given a parameter $x\in\R$ and a set $E\subseteq\R^2$ with $\dimH(E)\geq 1$, what can we say about the Hausdorff dimension of the vertical \emph{slice} $E_x=\{y:(x,y)\in E\}$? Without further information, we can only give the trivial upper bound,
	\begin{equation}\label{eq:slice}
		\dimH(E_x)\leq 1\,.
	\end{equation}
 	For instance, equality holds in~(\ref{eq:slice}) whenever $\{x\}\times[0,1]\subseteq E$. It would be more informative, then, to ask about the Hausdorff dimension of a \emph{random} vertical slice of $E$. The Marstrand slicing theorem tells us that if $E$ is a Borel set, then for Lebesgue almost every $x\in E$,
	\[\dimH(E_x)\leq \dimH(E_x)-1\,.\]
	Several more general results giving upper bounds on the Hausdorff dimension of the intersections of random transformations of restricted classes of sets have been proven, including theorems by Mattila~\cite{Matt84,Matt85,Matt95} and Kahane~\cite{Kaha86}; in particular, Falconer~\cite{Falc14} showed that when $E,F\subseteq\R^n$ are Borel sets,
	\begin{equation}
	\dimH(E\cap(F+z))\leq\max\{0,\dimH(E\times F)-n\}
	\end{equation}
	holds for Lebesgue almost every $z\in\R$. Using the point-to-set principle, N. Lutz showed that this inequality holds even when the Borel assumption is removed.

	\begin{theorem}[\cite{Lutz17}]\label{thm:int}
		For all $E,F\subseteq\R^n$, and for Lebesgue almost every $z\in\R^n$,
		\begin{equation*}
		\dimH(E\cap(F+z))\leq\max\{0,\dimH(E\times F)-n\}\,.
		\end{equation*}
	\end{theorem}

\section{Kakeya Sets and Generalized Furstenberg Sets}\label{sec:kgf}

A \emph{Kakeya set} in $\R^n$ is a set that contains unit-length line segments in all directions. That is, a set $E\subseteq\R^n$ such that for every direction $a\in S^{n-1}$ (the $(n-1)$-dimensional unit sphere in $\R^n$), there exists $b\in\R^{n}$ with $\{ax+b\mid x\in[0,1]\}\subseteq E$.

Besicovitch~\cite{Besi19,Besi28b} proved that Kakeya sets in $\R^n$ can have measure 0, and Davies~\cite{Davi71} proved that Kakeya sets in $\R^2$ must have Hausdorff dimension 2.

J. Lutz and N. Lutz gave computability theoretic proofs of both of these facts. They showed that the former corresponds to the existence of lines in all directions that contain no random points~\cite{LutLut15a}, and that the latter corresponds to the fact that for any random pair $(a,x)\in\R^2$, $\dim(x,ax+b)=2$ holds for all $b\in\R$~\cite{LutLut18}.

A set $E\subseteq\R^2$ is an \emph{$(\alpha,\beta)$-generalized Furstenberg set}, for parameters $\alpha,\beta\in[0,1]$, if $E$ contains $\alpha$-dimensional subsets of lines in all of a $\beta$-dimensional set of directions. That is, $E$ is an $(\alpha,\beta)$-generalized Furstenberg set if there is a set $J\subseteq S^1$ such that $\dimH(H)=\beta$ and, for every direction $a\in J$, there exist $b\in\R^2$ and $F_a\subseteq\R$ with $\dimH(F_a)=\alpha$ and $\{ax+b\mid x\in S_a\}\subseteq E$.

It is known that $(\alpha,\beta)$-generalized Furstenberg sets of Hausdorff dimension $\alpha+\frac{\alpha+\beta}{2}$ exist. Molter and Rela~\cite{MolRel12} gave a lower bound on the Hausdorff dimension of such sets:
\begin{equation}\label{eq:mr}
	\dimH(E)\geq\alpha+\max\left\{\frac{\beta}{2}, \alpha+\beta-1\right\}\,.
\end{equation}
Stull~\cite{Stul18} gave a new computability theoretic proof of~(\ref{eq:mr}), based on the point-to-principle. N. Lutz and Stull used the point-to-set principle to give a bound that improves on~(\ref{eq:mr}) whenever $\alpha,\beta<1$ and $\beta<2\alpha$.

\begin{theorem}[\cite{LutStu20}]\label{thm:furstenberg}
	For all $\alpha,\beta\in(0,1]$ and every set $E\in F_{\alpha\beta}$,
	\[\dimH(E)\geq \alpha+\min\{\beta,\alpha\}\,.\]
\end{theorem}

\section{Fractal Projections}\label{sec:proj}

In recent decades, Marstrand's projection theorem has become one of the most central results in fractal geometry~\cite{FaFrJi15}. It says that almost all orthogonal projections of a Borel set onto a line have the maximum possible dimension. More formally, letting $\proj_a$ denote orthogonal projection onto a line in direction $a$, Marstrand's projection theorem states that for all Borel $E\subseteq \R^2$ and Lebesgue almost every $a\in S^1$,
\begin{equation}\label{eq:proj}
	\dimH(\proj_a E)=\min\{1,\dimH(E)\}\,.
\end{equation}
Given Theorems~\ref{thm:prod} and~\ref{thm:int}, it is natural to hope that the point-to-set principle for Hausdorff dimension might allow us to remove the Borel assumption here as well. But Davies~\cite{Davi79}, assuming the continuum hypothesis, constructed a non-Borel set $E$ for which~(\ref{eq:proj}) does not hold. Nevertheless, N. Lutz and Stull used the point-to-set principles to prove the following.

\begin{theorem}[\cite{LutStu18}]
	Let $E\subseteq \mathbb{R}^2$ be any set such that $\dimH(E)=\dimP(E)$. Then for Lebesgue almost every $a\in S^1$,
	\[\dimH(\proj_a E)=\min\{1,\dimH (E)\}\,.\]
\end{theorem}

\begin{theorem}[\cite{LutStu18}]
	Let $E\subseteq \mathbb{R}^2$ be any set. Then for Lebesgue almost every $a\in S^1$,
	\[\dimP(\proj_a E)\geq\min\{1,\dimH (E)\}\,.\]
\end{theorem}

\section{Conclusion}\label{sec:conc}

As the preceding four sections show, the point-to-set principles have enabled the theory of computing to make significant advances in classical fractal geometry in a very short time. There is every indication that more such advances are on the near horizon.  But a scientist with Ker-I Ko's vision would already be asking about more distant horizons. What other areas of classical mathematical analysis can be advanced by analogous methods?  Are there intrinsic limits of such methods? We look forward to seeing the answers to these questions take shape.

\bibliography{wau}
\end{document}